\documentclass[twocolumn,showpacs,preprintnumbers,amsmath,amssymb]{revtex4}

\usepackage{graphicx}
\usepackage{dcolumn}
\usepackage{bm}
\usepackage{color,ulem}

\begin{document}

\preprint{}

\title{Anomalous enhancement of spin Hall conductivity in superconductor/normal metal junction}

\author{S. Hikino$^{1,2}$}
\author{S. Yunoki$^{1,2,3}$}%
\affiliation{%
$^{1}$Computational Condensed Matter Physics Laboratory, RIKEN ASI, Wako, Saitama 351-0198, Japan \\
$^{2}$CREST, Japan Science and Technology Agency (JST), Kawaguchi, Saitama 332-0012, Japan\\
$^{3}$Computational Materials Science Research Team, RIKEN AICS, Kobe, Hyogo 650-0047, Japan
} 

\date{\today}

\begin{abstract}
We propose a spin Hall device to induce a large spin Hall effect in a superconductor/normal 
metal (SN) junction. 
The side jump and skew scattering mechanisms are both taken into account 
to calculate the extrinsic spin Hall conductivity in the normal metal. 
We find that both contributions are 
anomalously enhanced when the voltage between the superconductor and the normal metal 
approaches to the superconducting gap. 
This enhancement is attributed to the resonant increase of the density of states 
in the normal metal at the Fermi level. 
Our results demonstrate a novel way to control and amplify the spin Hall conductivity by applying an external dc 
electric field, suggesting that a SN junction has a potential application for a spintronic device with a large 
spin Hall effect. 
\end{abstract}

\maketitle

How to generate and manipulate spin current is one of the central issues in the research field of 
spintronics~\cite{zutic-rev, maekawa-book}. In particular, the ability to control spin current 
by an external electric field is essential because the electric field can control the flow of 
electrons in nanometer scale devices. In this regard, an interaction between the spin and 
orbital motion of electrons [spin-orbit interaction (SOI)] is an important ingredient. 
The SOI induces the novel phenomenon called spin Hall effect (SHE), 
where a charge current induces spin dependent motion of electrons, flowing perpendicular 
to the charge current and in the opposite directions for up- and down-spin electrons, 
and thus the spins are accumulated at the edge of the sample. The SHE has been recognized as a key effect to 
convert the charge current into the spin current and vice 
versa~\cite{karplus,d'yakonov, hirsh, zhang, nagaosa-rev}. 

The SHE was first predicted theoretically decades ago, and now it is well accepted that there are 
two types of SHE, the one caused by the SOI of a host metal ({\it intrinsic} SHE)~\cite{karplus} 
and the other caused by the SOI of nonmagnetic guest impurities (most often heavy elements) 
in a host metal ({\it extrinsic} SHE)~\cite{d'yakonov}. 
For the extrinsic SHE, there are two contributions, skew scattering and side jump. 
The skew scattering results from the impurity scattering via the SOI~\cite{smit}, 
whereas the side jump is due to the anomalous velocity induced by the SOI~\cite{berger}.  
The first experimental observation of the extrinsic SHE has been reported by Kato $et$ $al$, 
who have detected the spin accumulation induced by the extrinsic SHE in GaAs systems~\cite{kato,wunderlich}. 
Their work has stimulated extensive theoretical as well as experimental studies for the extrinsic SHE in various materials 
with different experimental setups~\cite{ex-she, theory-she, tse,stakahashi, seki, koong, guo-gu, gradhand}.  

One of the important current issues is to find a way to obtain a large SHE~\cite{seki, koong, guo-gu, gradhand}. 
A large spin Hall conductivity (SHC) has an ability to generate the large spin current. 
The SHC in turn depends sensitively on the SOI as well as the impurity scattering in a host material. 
Very often, a larger SHE has been observed experimentally in impurity doped (extrinsic) systems rather than in 
impurity free (intrinsic) systems~\cite{kato, ex-she,morota}. This is, in fact, consistent with theoretical 
calculations~\cite{seki, koong, guo-gu, gradhand}. 
However, the SHE observed is still small, requiring sensitive experimental measurements to detect 
the effect, in host materials such as light element metals (Al or Cu) and semiconductors~\cite{kato, ex-she}. 
Therefore, finding an alternative way to further increase the SHE is highly desirable, which certainly helps to achieve a variety of 
SHE-based spintronic devices in the future. This is precisely the main purpose of this paper. 

\begin{figure}[!t]
\begin{center}
\vspace{6mm}
\includegraphics[width=5.5cm]{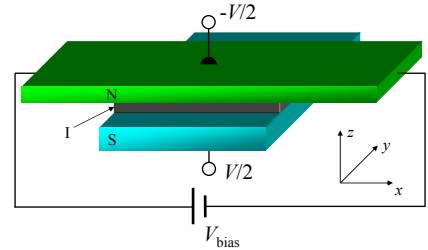}
\caption{(Color online) Schematic configuration of a SN junction proposed to induce a large SHE. 
$V_{\rm bias}$ is an applied dc voltage at the opposite edges of the N. $V$ is a 
voltage applied between S and N. An insulating barrier (I) is inserted between S and N. 
}
\label{sn-gm}
\end{center}
\end{figure}

In this paper, we propose a simple superconductor/normal metal (SN) junction in which a large SHE is induced. 
Taking into account both contributions of the side jump and skew scattering mechanisms in low impurity 
concentrations, we show that the extrinsic SHC in the normal metal is anomalously enhanced when the 
voltage between the superconductor (S) and the normal metal (N) approaches to the superconducting gap. 
This enhancement is attributed to the resonant increase of the density of states in the N at the Fermi level. 
Our results demonstrate that the SHC can be controlled and amplified by using the dc voltage, 
suggesting that a SN junction has a potential application for spintronic devices with a large SHE. 

The system considered is an $s$-wave SN junction as depicted in Fig.$~$\ref{sn-gm}. 
An insulating barrier is inserted between S and N to suppress the proximity effect. 
In this setup, the thickness of the N is considered thin enough to treat the N as a two dimensional N. 
A dc bias voltage $V_{\rm bias}$ is applied in $x$-direction to flow electrons in the N. 
The chemical potential difference between S and N is adjusted by a dc voltage $V$ applied in $z$-direction~\cite{note}. 
The system is thus described by the Hamiltonian $H = H_{\rm S} + H_{\rm em}^{\rm N} + H_{\rm N} + H_{\rm T}$. 
Here $H_{\rm S}$ is the Bardeen-Copper-Schrieffer (BCS) Hamiltonian with an $s$-wave superconducting gap. 
$H_{\rm em}^{N}$ represents the interaction with the applied dc bias voltage $V_{\rm bias}$: 
$H_{\rm em}^{\rm N} = -\int d^{2} r {\bm j} (\bm{r},t)\cdot {\bm A}(t)$ 
where ${\bm j} (\bm{r},t)$ and ${\bm A}(t)$ are a current operator (defined below) and a vector potential, respectively. 
The gauge is set to satisfy ${\bm E} = -\partial _{t} {\bm A (t)}$ with a spatially 
uniform electric field [${\bm A}(t) = (-E_x t, 0,0)$]. 
The N is described by $H_{\rm N} = \sum_{\sigma} \int d^{2} r c_{\sigma}^{\dagger} (\bm{r}) \left(-\frac{ \hbar^{2} }{ 2m } \nabla^{2} - \mu_{\rm F}\right)c_{\sigma}(\bm{r})
 + H_{\rm {imp}} + H_{\rm SOI}$, 
where $c_{\sigma}(\bm{r})$ is an annihilation operator of electron with spin $\sigma$ at position $\bm r$. 
$m$ and $\mu_{\rm F}$ are mass of electron and the Fermi level, respectively. 
The terms $H_{\rm imp}$ and $H_{\rm SOI}$ describe a nonmagnetic impurity scattering and the SOI, respectively,  
\begin{eqnarray}
H_{\rm{imp}} &=& 
		\sum_{\sigma} \int d^{2} r
		u({\bm r})
		c_{\sigma}^{\dagger}({\bm r}) c_{\sigma}({\bm r})
,\nonumber \\
H_{\rm {SOI}} &=& 
		-i \lambda_{\rm SO}
		\sum_{\alpha, \beta}
		\int d^{2} r
		c_{\alpha}^{\dagger} ({\bm r}) 
		\left[
		\nabla u({\bm r})\cdot \nabla \times {\bm \sigma}_{\alpha,\beta} 
		\right]
		c_{\beta} (\bm r), \nonumber
\end{eqnarray}
where $u(\bm r)=u_{0} \sum_{i} \delta ({\bm r} - {\bm R}_{i})$ is an impurity potential with the 
strength $u_{0}$ locating at $\bm R_{i}$ in the N. 
$\lambda_{\rm SO}$ is the SOI coupling and ${\bm \sigma}_{\alpha, \beta}$ are the Pauli matrices. 
For the tunneling of electrons between S and N, we adopt the tunneling Hamiltonian $H_{\rm T}$ described by 
\begin{eqnarray}
H_{\rm T} &=&
		\sum_{\sigma} \int_{ {\bm r}\in {\rm N}, {\bm r'}\in {\rm S} } d^{2} r d^{3} r'
		T_{\bm r, \bm r'} e^{i \frac{eV}{\hbar} t}
		c_{\sigma}^{\dagger} (\bm r) d_{\sigma} (\bm r')
		+ {\rm h.c.}
. \nonumber
\end{eqnarray}
Here, $d_{\sigma}(\bm r)$ is an annihilation operator of electron in the S and the tunneling matrix element 
$T_{{\bm r},{\bm r'}}$ is non zero only at the SN boundary, i.e.,  
$T_{\bm r, \bm r'} = T_{0} \delta(\bm r - \bm r'_{\parallel }) \delta(z')$ with $\bm r'_\parallel $ = $(x', y',0)$. 
Finally, the voltage $V$ between S and N is described by the exponential factor 
$e^{ieVt/\hbar}$ in $H_{\rm T}$. 

\begin{figure}[t]
\begin{center}
\includegraphics[width=6.5cm]{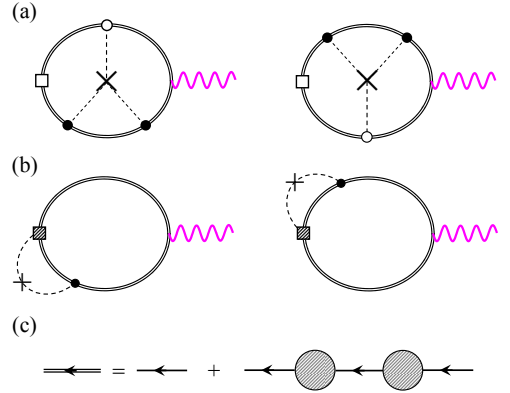}
\caption{(Color online) The lowest order diagrams for the skew scattering (a) and side jump (b) contributions 
to the spin Hall conductivity. Wavy lines denote the vector potential.
Open squares (shadow squares) are vertices of normal velocity (anomalous velocity).  
Solid (open) circles represent scatterings via the nonmagnetic impurity (spin-orbit interaction) described 
by $H_{\rm imp}$ ($H_{\rm SOI}$). Crosses indicate impurities. 
(c) The electron Green's function in the N up to the second order of the tunneling 
matrix element. 
Here large solid circles indicate the tunneling matrix element. 
}
\label{diagram-shc}
\end{center}
\end{figure}

To evaluate the extrinsic SHC within the linear response theory, 
first we consider the statistical average of the following two current operators in the $y$-direction~\cite{hosono} 
\begin{eqnarray}
j_{y, \sigma}^{ \rm N } (\bm{r}, t) &=& 
		-i \frac{e\hbar}{m S_A}
		\sum_{\bm{k},\bm{q}} e^{-i \bm{q} \cdot \bm{r} } k_{y}
		\left<
		G_{\bm k+\frac{\bm q}{2}, \bm k-\frac{\bm q}{2}, \sigma,\sigma}^{-+} (t,t) 
		\right>_{\rm i},
\nonumber \\
%
%
j_{y,\alpha}^{\rm SO} (\bm{r},t) &=& 
		i \frac{e\lambda_{ \rm SO }}{ \hbar S_A }
		\sum_{ \bm{k}, \bm{q}, \beta}
		e^{-i \bm{q} \cdot \bm{r} } \nonumber \\
		&&\times
		\left<
		\left[\nabla U (\bm r)
		\times
		\bm{\sigma}_{\beta \alpha}
		\right]_{y}
		G_{\bm k+\frac{\bm q}{2}, \bm k-\frac{\bm q}{2}, \alpha, \beta}^{-+} (t,t)
		\right>_{\rm i},
		\nonumber
\end{eqnarray}
where $G_{\bm k', \bm k, \sigma', \sigma}^{-+} (t',t) = i\left< c_{\bm k, \sigma}^{\dagger}(t) c_{\bm k', \sigma'}(t') \right>$ 
is a lesser Green's function, $\bm k$ and ${\bm k}'$ are wave numbers of electrons, and $S_A$ is the area of the junction. 
$\langle\cdots \rangle_{\rm i}$ represents the impurity average. 
The lesser Green's function is derived from the contour Green's function, 
$G_{\bm k, \bm k', \sigma, \sigma'}(t,t')=-i\left<T_{c} c_{\bm k, \sigma}(t)c_{\bm k', \sigma'}^{\dagger}(t') \right>$, 
where $\langle\cdots \rangle$ denotes the quantum statistical average at zero temperature and 
$T_{c}$ is a contour ordering operator~\cite{book-keldysh1}. 
$j_{y, \sigma}^{ \rm N } (\bm{r}, t)$ is the normal current, from which the skew scattering contribution is obtained, 
while $j_{y,\alpha}^{\rm SO} ({\bm r},t)$ is the anomalous current originated from the SOI term, from which 
the side jump contribution is obtained. 

The SHC is obtained from 
$j_{y,\sigma}^{\rm N}({\bm r},t) = \sigma \sigma_{xy}^{\rm{ SS} } E_x$ and 
$j_{y,\sigma}^{\rm{SO}}({\bm r},t) = \sigma \sigma_{xy}^{\rm{ SJ} } E_x$, 
where $\sigma_{xy}^{\rm SS}$ ($\sigma_{xy}^{\rm SJ}$) is the skew scattering (side jump) 
contribution to the SHC, and $\sigma=+1$ ($-1$) for up (down) electrons.  
The terms $H_{\rm imp}$ and $H_{\rm SOI}$ are treated 
within a perturbation theory keeping the lowest order contributions, denoted by Feynman diagrams shown in 
Fig.~\ref{diagram-shc} (a) and (b). 
This approximation is valid in a low impurity concentration $n_{\rm imp}$. 
The SHC due to the skew scattering and side jump mechanisms is then summarized as 
\begin{eqnarray}
\sigma_{xy}^{\rm SS} &=&
			i\frac{ e^{2} \lambda_{\rm SO} u_{0}^{3} n_{\rm imp} }{ 2\pi m^{2} \hbar^{2} S_A^{3} }
			\sum_{{\bm k}, {\bm k'}, {\bm k''}}   
			k_{y}^{2} G_{\bm k}^{\rm R} G_{\bm k}^{\rm A} \nonumber \\ 
			&\times& {k'}_x^2 G_{\bm k'}^R G_{\bm k'}^{\rm A}
			\left(
			G_{\bm k''}^{\rm R} - G_{\bm k''}^{\rm A}
			\right) 
\label{ssqp}, \\
\sigma_{xy}^{\rm SJ} &=& 
		\frac{e^{2} \lambda_{\rm SO} u_{0}^{2}n_{\rm imp} }{2\pi \hbar^{2}S_A^{2}m}
		\sum_{{\bm k},{\bm k'}} k_{x}^{2} 
		G_{\bm k}^{\rm R} G_{\bm k}^{\rm A} 
		\left(G_{\bm k'}^{\rm R} - G_{\bm k'}^{\rm A} \right) 
\label{sjqp}, 
\end{eqnarray}
where $G_{\bm k}^{\rm R}$ ($G_{\bm k}^{\rm A}$) is the retarded (advanced) Green's function with zero frequency. 

\begin{figure}[!t]
\begin{center}
\vspace{1mm}
\includegraphics[width=6cm]{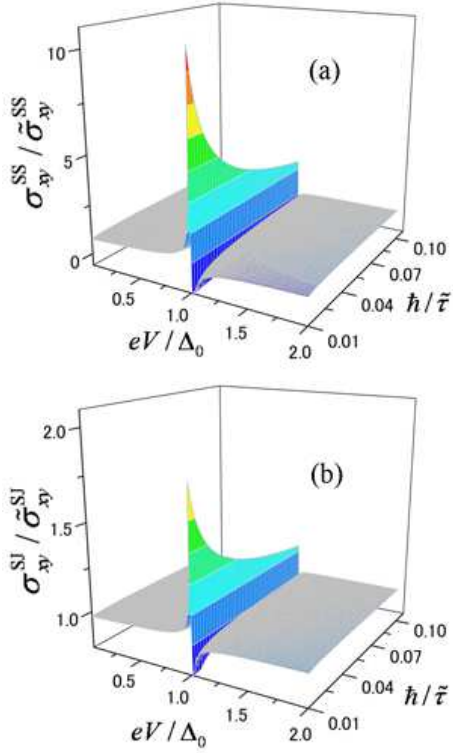}
\caption{(Color online) Spin Hall conductivity for the skew scattering (a) and side jump (b) contributions as functions of 
the voltage $V$ and the inverse relaxation time $\tau^{-1}$. 
Here ${\tilde\tau}=\tau\Delta_0$, ${\tilde{\sigma}}_{xy}^{\rm SS}$ (${\tilde{\sigma}}_{xy}^{\rm SJ}$) is the skew scattering 
(side jump) contribution in the bulk N,  
and a dimensionless parameter $g_{0}=\frac{T_{0}^{2}\sqrt{m}}{\hbar \pi (2\Delta_{0})^{3/2}}$ 
is set to be $2\times10^{-5}$~\cite{note2}. 
}
\label{shc}
\end{center}
\end{figure}

To calculate $G_{\bm k}^{\rm R(A)}$ for the N in the SN junction, we keep the lowest contribution of the tunneling matrix 
element $T_0$, 
which is indicated by Dyson's equation shown in Fig.$~$\ref{diagram-shc} (c), and we obtain 
\begin{equation}
G_{\bm k}^{\rm R(A)} = g_{\bm k}^{\rm R(A)} + \frac{T_{0}^{2}}{\hbar^{2}} 
		g_{\bm k}^{\rm R(A)} g_{{\rm S},\bm k}^{\rm R(A)}(eV) g_{\bm k}^{\rm R(A)} 
\label{gf2}, 
\end{equation}
where $ g_{\bm k}^{\rm R(A)} = \hbar/({-\xi \pm i\hbar/2\tau})$
is the retarded (advanced) Green's function in the N. 
$\xi = \hbar^{2}k^{2}/2m-\mu_{\rm F}$ is the kinetic energy of electron 
and $\tau$ is the relaxation time due to the nonmagnetic impurity scattering within the Born approximation. 
$g_{{\rm S},\bm k}^{\rm R(A)}(eV)$ is the diagonal part of the retarded (advanced) Green's function in the S given by 
\begin{equation}
g_{{\rm S},\bm k}^{\rm R(A)}(eV) = -i\frac{m}{2\hbar}
		\left[
		\frac{ eV}{i\Omega_{\rm R(A)}} \left( \frac{1}{p_{\uparrow }} + \frac{1}{p_{\downarrow }} \right)
		+\frac{1}{p_{\uparrow }} - \frac{1}{p_{\downarrow }}
		\right]
\label{gfs}, 
\end{equation}
which satisfies the Gorkov's equation. 
Here, $p_{\uparrow (\downarrow) } = \sqrt{2m( -\xi \pm i\Omega_{\rm R (A)} )/\hbar^{2}}$ and 
$i \Omega_{\rm R (A)} = \sqrt{(eV\pm i\eta)^{2} - \Delta_0^{2} }$. 
$\eta$ is the inelastic scattering rate in the S~\cite{rodero} and $\Delta_{0}$ is the superconducting gap at zero temperature. 
Substituting Eqs.$~$(\ref{gf2}) and (\ref{gfs}) into Eqs.$~$(\ref{ssqp}) and (\ref{sjqp}), 
we obtain for the SHC 
\begin{widetext}
\begin{eqnarray}
\frac{\sigma_{xy}^{\rm SS}}{{\tilde{\sigma}}_{xy}^{\rm SS}} &=&
		1
		+
		\frac{4 T_{0}^{2}}{\hbar^{3} \tau}
		\int_{-\infty}^{\infty}d\xi
		\Re
		\left[
		(g_{\bm k}^{\rm R})^{2} 
		g_{\bm k}^{\rm A}
		g_{{\rm S}, {\bm k}}^{\rm R}(eV)
		\right]
		-
		\frac{2T_{0}^{2}}{\hbar^{3}}
		\int_{-\infty}^{\infty }d\xi
		\Im
		\left[
		\left(
		g_{\bm k}^{\rm R} 
		\right)^{2}
		g_{{\rm S}, {\bm k}}^{\rm R}(eV)
		\right]
		, \nonumber \\
\frac{\sigma_{xy}^{\rm SJ}}{{\tilde{\sigma}}_{xy}^{\rm SJ}} &=&
		1
		+
		\frac{T_{0}^{2}}{\hbar^{3} \pi \tau}
		\int_{-\infty}^{\infty} d\xi
		\mid g_{\bm k}^{\rm R} \mid^{2} 
		\Re
		\left[
		g_{\bm k}^{\rm R}
		g_{{\rm S}, {\bm k}}^{\rm R}(eV)
		\right]
		-
		\frac{T_{0}^{2}}{\hbar^{3} \pi}
	    \int_{-\infty}^{\infty} d\xi
		\Im
		\left[
		\left(
		g_{\bm k}^{\rm R} 
		\right)^{2}
		g_{{\rm S}, {\bm k}}^{\rm R}(eV)		
		\right], \nonumber
\end{eqnarray}
\end{widetext}
where ${\tilde{\sigma}}_{xy}^{\rm SS}$ (${\tilde{\sigma}}_{xy}^{\rm SJ}$) is the skew scattering 
(side jump) contribution to the SHC in the bulk N with $T_0=0$. 


Let us now evaluate numerically the SHC for the two contributions derived above. 
Here we take $\eta/\Delta_{0}=1\times10^{-3}$~\cite{inelastic-scatt}. 
Fig.~\ref{shc} (a) shows the skew scattering contribution to the SHC ($\sigma_{xy}^{\rm SS}$), 
normalized by the SHC for the bulk N (${{\tilde{\sigma}}_{xy}^{\rm SS}}$), as functions of 
$V$ and the inverse relaxation time $\tau^{-1}$. 
From Fig.~\ref{shc} (a), it is observed that $\sigma_{xy}^{\rm SS}$ is almost the same as that of the bulk system 
when $eV$ deviates from $\Delta_{0}$. 
However, when $eV$ approaches to $\Delta_{0}$, 
$\sigma_{xy}^{\rm SS}$ becomes anomalously enhanced. 
Moreover, it is seen that $\sigma_{xy}^{\rm SS}$ monotonically increases with $\tau$. 
Fig.~\ref{shc} (b) shows the side jump contribution to the SHC ($\sigma_{xy}^{\rm SJ}$), which 
exhibits the similar characteristic behavior, i.e., large enhancement of 
$\sigma_{xy}^{\rm SJ}$ for $eV$ close to $\Delta_0$, although the enhancement factor for $\sigma_{xy}^{\rm SJ}$ 
appears smaller than that for $\sigma_{xy}^{\rm SS}$. 
These results clearly demonstrate that $\sigma_{xy}^{\rm SS}$ and $\sigma_{xy}^{\rm SJ}$ can be significantly amplified 
by tunning $V$ between S and N in the SN junction. 

\begin{figure}[!t]
\begin{center}
\vspace{1mm}
\includegraphics[width=6cm]{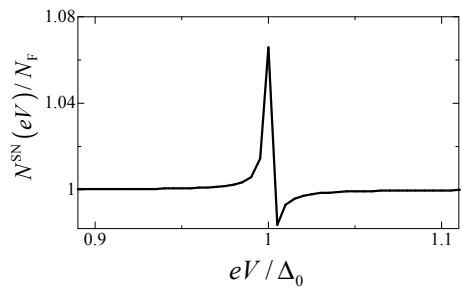}
\caption{The DOS [$N^{\rm SN}(eV)$] at the Fermi level for the N as a function of voltage $V$ in the SN junction 
at zero temperature. 
Here, we take $g_0=2\times10^{-5}$~\cite{note2} and 
$\hbar/\tilde \tau=10^{-2}$. $N_{\rm F}$ is the DOS at the Fermi level for the bulk N. 
}
\label{dos}
\end{center}
\end{figure}

Next, we shall elucidate the origin of this enhancement. 
To this end, it is important to notice that the extrinsic SHC is related to the electron density 
of states (DOS) at the Fermi level. 
The DOS at the Fermi level in the N for the SN junction considered here is obtained 
by taking the imaginary part of the retarded Green's function 
($N^{\rm SN} = -\frac{1}{\hbar \pi} \Im [G^{\rm R}]$), 
which is calculated by considering the same diagram shown in Fig.~\ref{diagram-shc} (c). 
As shown in Fig.~\ref{dos}, the DOS for the N shows a sharp peak for $eV$ close to $\Delta_0$. 
This resonant increase of the DOS is simply due to tunneling of normal electrons to the S, thus reflecting the DOS of the S 
which is singularly large for energy around $\Delta_{0}$~\cite{note-tau}. 

%
Let us now estimate the value of the relaxation time $\tau$ since this quantity determines rather sensitively 
the value of the SHC as already shown in Fig.~\ref{shc}. 
When the temperature is much lower than a typical phonon frequency, the relaxation time $\tau$ in the N 
is mainly determined by the elastic impurity scattering. In this temperature region, the value of $\tau$ for the N 
such as Cu and $n$-GaAs is roughly estimated to be in a range of 10--100~ps~\cite{tse,scale}. 
For instance, $\hbar/\tau\Delta_0\sim0.01$ for $\Delta_{0}\simeq 1$~meV and $\tau\simeq 100$~ps, 
and  thus 
the enhancement factor of the SHC due to the skew scattering (side jump) is as large as about 10 (less than 2). Therefore, 
we expect that the anomalous enhancement of the SHC should be easily observed experimentally. 

Finally, several notes are in order. First, we have considered the extrinsic SHE in this paper. 
We expect the similar enhancement for the intrinsic SHE in a SN junction. 
Second, it is well known that the tunneling matrix element $T_0$ in the SN junction is 
inversely proportional to the resistance through the junction~\cite{dittrich-book}. 
Here, only the lowest contribution of $T_0$ is considered, assuming that the system studied 
has a comparatively high interface resistance~\cite{note2}. 
In such a junction, the proximity effect and the higher order contributions of $T_0$ are safely neglected. 
However, for more quantitative analysis, details of the interface structure have to be taken into account.  
Third, the spatial variation of the superconducting order parameter in the S near the interface, which has not 
been considered in this study, becomes important for a small tunnel barrier 
(or even for a junction with a metallic interface between S and N). 
Such variation of the order parameter would affect the SHC, which remains to be studied in the future. 

In summary, we have proposed a simple SN junction to induce a large extrinsic SHE. 
The side jump and skew scattering contributions have been taken into account to calculate 
the SHC in low impurity concentrations. 
We found that both contributions are anomalously enhanced 
when the voltage between S and N is adjusted close to $\Delta_{0}$. 
This enhancement is attributed to the resonant increase of the DOS in the N at the Fermi level. 
We believe that this enhancement of the SHC is large enough to be observed 
experimentally~\cite{culcer}. 
Our results demonstrate that the SHC can be controlled and amplified by using a dc electric field, 
suggesting that a SN junction has a potential application for a spintronic device with a large SHE. 

The authors thank S. Takahashi and Y. Niimi for valuable discussions and comments. 


\begin{thebibliography}{}
\bibitem{note}
Note that the chemical potential of the N is equal to that of the S when $V=0$.
\bibitem{note-tau}
The apparent difference in the enhancement factors for SHC and DOS is simply because the enhancement factor for 
SHC contains $\tau$ while that for DOS does not.
\end{thebibliography}
\end{document}